\documentclass[onecolumn]{emulateapj}

\shorttitle{On the lack of dark matter in  NGC 1052–DF2 galaxy}
\shortauthors{C. E. Navia}

\usepackage{amsmath}

\begin{document}


\title
{
 On the Generalized Faber-Jackson relation for NGC 1052–DF2 galaxy
}


\author{Carlos. E. Navia }
\affil{Instituto de F\'{i}sica, Universidade Federal Fluminense, 24210-346, Niter\'{o}i, RJ, Brazil}


\altaffiltext{1}{E-mail address:navia@if.uff.br}


\begin{abstract}
The conjectures of Hawking-Bekenstein of that gravity and temperature are closely related, was the beginning of the formulation of thermodynamic models of gravity. Using one of these models, we show that the thermal bath temperature of NGC1052-DF2 dwarf galaxy is about 6.5 K higher than the temperature of the thermal bath of the local dwarf galaxies. We claim that this temperature difference is responsible
by the behavior of the NGC1052-DF2 galaxy  to be apparently a galaxy with lack dark matter. This difference arises due to that the host galaxy of NGC1052-DF2, the elliptic galaxy NGC1052 is a LINER-type active galactic nucleus, with signals the intense starburst activity in the galaxy's center.
Debye Gravitational Theory (DGT), allows obtaining a Generalized Faber-Jackson relation to described the dwarf galaxies  and obtaining
 the velocity dispersion as a function of thermal bath temperature.
 DGT predict a velocity dispersion to NGC1052-DF2 dwarf galaxy in agreement with the very-low values reported in the literature. Also, DGT predicts that several, or maybe all dwarf galaxies from NGC1052 must have the same behavior than NGC1052-DF2 because they are immersed in thermal baths almost with the same temperature, this behavior already was observed in a second galaxy, the NGC1052-DF4 dwarf galaxy. Also, DGT predicts the same behavior,  in all satellite galaxies orbiting galaxies with active nuclei.   

\end{abstract}





\section{Introduction}

Just as the Milky Way has a large number of satellite galaxies (dwarf galaxies) orbiting it. The giant elliptical galaxy NGC 1052 located at a distance of 20 Mpc from the solar system,  in the Cetus constellation also has about 22 satellite galaxies. One of these, the ultra diffusive galaxy named NGC 1052-DF2 shows an anomaly, recent observations by \cite{vand18} show that this galaxy has a very low-velocity dispersion only of 7.8 km s$^{-1}$ and that corresponds to a stellar mass of $ 2 \times 10^8$ solar masses. 
However, the total mass within a radius of 7.6 kpc is estimated to be less than $ 3.4 \times 10^8$ solar masses. That is, very close to the stellar mass, i.e., this galaxy is
compatible with a mass near zero for its dark matter halo. Thus we have the `` galaxy lacking dark matter ''.

We believe that this anomaly is a problem for models within the $\Lambda$CDM paradigm that invoke dark matter, 
understand how and why dark matter in a galaxy is diluted or separated from the luminous mass is a challenge. However, according to  Sokol\footnote{\url{https://www.quantamagazine.org/a-victory-for-dark-matter-in-a-galaxy-without-any-20180328/}} most delicate situation is for models that mimic dark matter, modifying the gravity, in this case, the modification would have to be always on.
Sokol among others claim that theories that challenge dark matter’s existence will need to explain away the new claim about galaxy NGC 1052–DF2 to survive. 

So far already at least three papers are reporting the velocity dispersion for NGC1052-NF2 within the MOND
paradigm \citep{krou18,mull19,fama18}, with values from 13.4 to 14.0 km s$^{-1}$.
Within the paradigm $\Lambda$CDM, the separation between dark matter and the luminous matter would be linked to tidal effects \citep{ogiy18}. However, so far there is no a clear causality that allows the introduction of feedback within the simulations to explain the anomaly, suggesting that the lack dark matter in the GCN1052-DF2 galaxy comes from the uncertainties in the mass estimates, that is substantially underestimated \citep{lapo18}.

Indeed, there are others measurements to the internal velocity dispersion of NGC1052-DF2, such using the deep MUSE
IFU observations (Emsellem et al. 2018) reporting a value of
$\sigma_{stellar} = 16.3\pm 5$ km s$^{-1}$, and a more smaller value when the dispersion is obtained from the globular 
clusters,  giving a value of $\sigma_{gc}=10.5^{+4.0}_{-2.2} $ km s$^{-1}$, as well as, through an independent analysis 
\citep{mart18}, the mean observed biweight for NGC1052-DF2 comes out as 
$\sigma_{obs,bi}= 14.3\pm 3.5$. km s$^{-1}$.
The above results show that NGC 1052-DF2 galaxy is still compatible with expectations from both, MOND and
$\Lambda$CDM  paradigms \citep{mart18}.

However, is claim by \cite{dani19} a new a more precise measurement, from high-resolution spectroscopy, 
using the light of NGC1052-DF2 obtained with the Keck Cosmic Web Imager (KCWI), with an instrumental resolution of  
12 km s$^{-1}$. Reporting a stellar velocity dispersion of 
$\sigma_{stars} = 8.4 \pm 2.1$ km s$^{-1}$, consistent with the dispersion that was derived from the globular clusters.

Also, almost simultaneously \citep{vand19} reports that this anomaly, lack of dark matter in galaxies,  is not restricted to the galaxy NGC1052-DF2. Of the 22 satellite galaxies that have the elliptical giant NGC1052, 
it seems that the diffuse dwarf galaxy NGC1052-DF4 also has a very low stellar velocity dispersion, $v_{stars}\sim7.$ kms$^{-1}$. According to \cite{vand19} this new observation of very low-velocity dispersion, in the NGC1052-DF4 galaxy, further, constrain models that invoke dark matter, and is in tension with recent predictions from alternative gravity.

In this paper, we try an alternative description of the NGC 1052–DF2 galaxy from DGT \citep{navi17}.
 DGT is a thermodynamic gravitational theory, it incorporates the structure of Debye theory of specific heat of solids at low temperatures, within the Entropic Gravitational Theory \citep{verl11}.  This framework is useful because it allows obtaining dynamic equations as a function of the temperature. DGT is very well successfully describing the galaxies rotation curves in a wide range of redshift \citep{navi17,navi18a}, the description of the local dwarf galaxies \citep{navi18b}, and the galaxy clusters \citep{navi19},  in all cases, DGT previsions are in agreement with the observations. 
 
\section{Method} 

 The observations of the galaxies NGC 1052–DF2 and NGC1052-DF4 as galaxies without an apparent dark matter, for us are in the same trail than the observations from ESO's Very Large Telescope (VLT) \citep{lang17}. Indicating that massive galaxies, during the peak epoch of galaxy formation, at redshift above 0.5 have declining rotation curves, below of the Keplerian one, this means that they were dominated by baryonic or ``normal'' matter.  Or in other words, they are ``galaxies lacking dark matter''. 
 In both cases, we show that the two systems: the NGC1052-DF2 galaxy and the VLT galaxies at high redshift (as described in \cite{navi18a}) are within thermal baths with temperatures at least $\sim 4.0$ degrees above 2.73 K. Under this condition and according to DGT the motion of the systems is more proximate to the Newtonian-regime than the Mondian-regime.

Let's begin with the Generalized Faber-Jackson relation from DGT, the same used to describe the galaxy cluster relations \citep{navi19}
\begin{equation}
M=\frac{r^{1-\alpha}}{Ga_0^{\alpha}} \sigma^{2\alpha+2},
\label{Eq:mass}
\end{equation}
where $a_0=1.2 \times 10^{-10}$ m s$^{-2}$ is the acceleration scale and the index value 
$\alpha$ is determined by the thermal bath temperature in which the system is immersed. 
The Eq~\ref{Eq:mass} has two asymptotic cases
\[ \alpha =
  \begin{cases}
    0      & \quad \text{then } M=(r/G)\sigma^2       \text{ (Newton-regime $T>13$\; K)}\\
    1      & \quad \text{then } M=(1/Ga_0) \sigma^4 \text{ (deep-MOND-regime $T\sim 2.7$\; K)}.\\
  \end{cases}
\]

\begin{figure}
\vspace*{-0.0cm}
\hspace*{0.0cm}
\centering
\includegraphics[width=13.0cm]{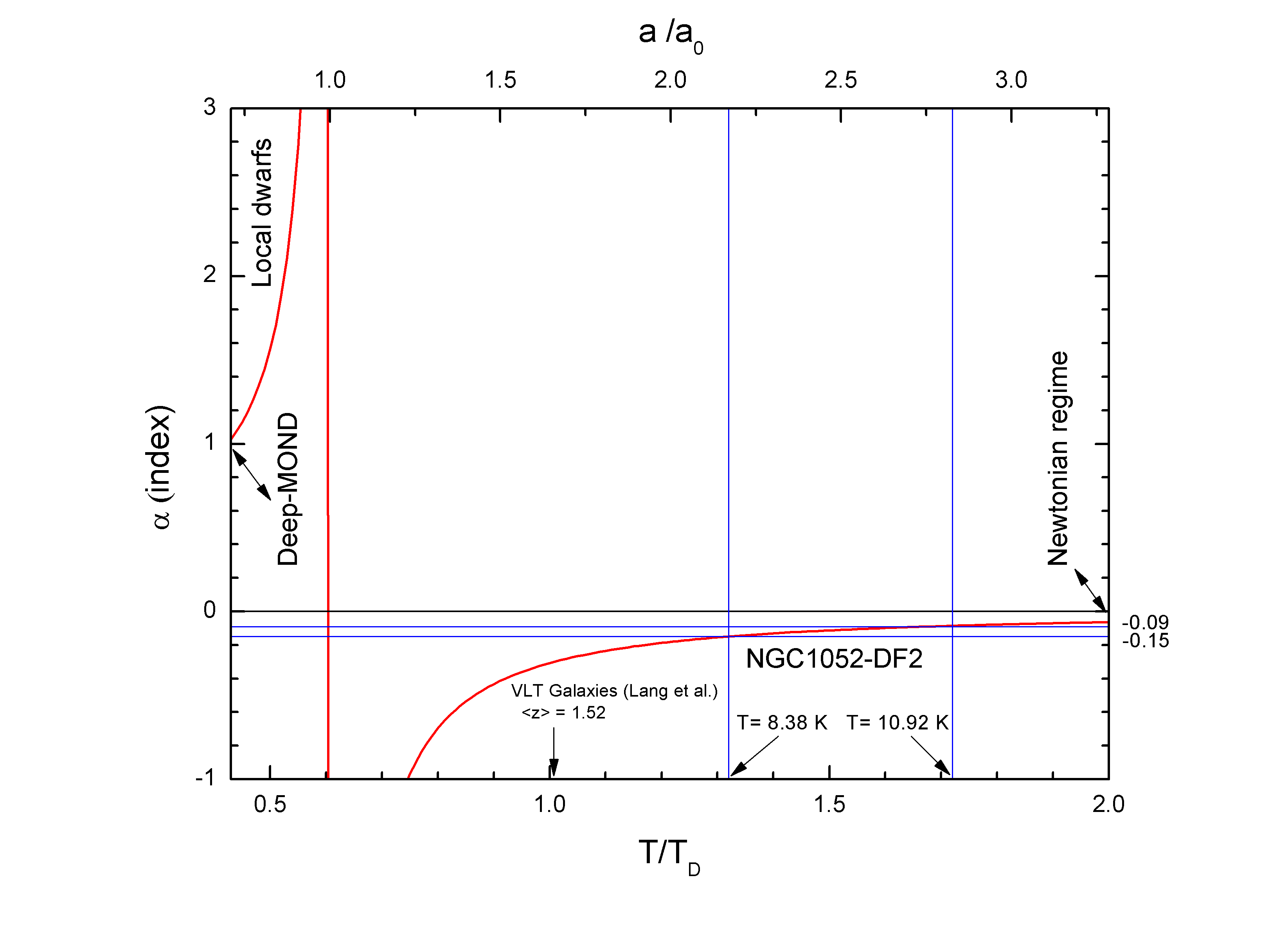}
\vspace*{-0.0cm}
\caption{Dependence of the $\alpha$ index with the temperature (acceleration) (red curves), according to Eq.~\ref{Eq:alpha_a}, with $T_D=6.35$ K. The two vertical blue lines indicate the interval of temperature for the thermal bath of  NGC1052-DF2 galaxy, obtained through an analysis that is summarized in Fig.~\ref{m_r}, see text.
}
\vspace*{0.0cm}
\label{alpha_a}
\end{figure}

In general, the index $\alpha$ satisfy the relation
 \begin{equation}
 \alpha = \frac{\log \mathcal{D}_1\left(\frac{a_0}{a}\right)}{\log \frac{a}{a_0}},
 \label{Eq:alpha_a}
 \end{equation}
 where $\mathcal{D}_1(x)$ is the first order Debye function \citep{navi17,navi19}.
In DGT there is a specific bond between the acceleration and temperature, give as
$a/a_0 = ¨(\pi^2/6)T/T_D$, where $T_D=6.35$ K, is the Debye temperature \citep{navi18a,navi18b}.
Fig.~\ref{alpha_a} shows the index $\alpha$ as a function of temperature
(bottom horizontal axis) and as a function of acceleration (top horizontal axis), according to Eq.~\ref{Eq:alpha_a}. 

We have shown in \cite{navi18b}, that DGT gives a good description of local dwarf galaxies,
considering that nearby isolated galaxies (z $\sim$ 0), such as the Milky-Way are immersed within a thermal bath at 2.73 K. 
However, the dwarf galaxies orbiting these galaxies are also subject to additional radiation from their hosts, so they are within a thermal bath slightly warmer, of up to one degree above 2.73 K. This thermal effect can explain several properties of the local dwarf galaxies. One degree above 2.73 K means in DGT a $\alpha > 0$ index above one and up to $\alpha \sim 9.0$. This region is located in the upper left of  Fig.~\ref{alpha_a}.

As already indicated, the VLT galaxies with an average redshift of $<z>=1.52$ \citep{lang17} (and means in DGT a bath temperature of $T=2.73(1+z)= 6.88$ K)   have declining rotation curves, an indication that they are ``galaxies lacking dark matter'' and they are described in DGT by in negative index  $\alpha=-0.26$
\citep{navi18a}. It is expected that the $\alpha$ index, to describe NGC1052-DF2 galaxy, also must be negative. 

\section{Results}

\begin{figure}
\vspace*{-0.0cm}
\hspace*{0.0cm}
\centering
\includegraphics[width=13.0cm]{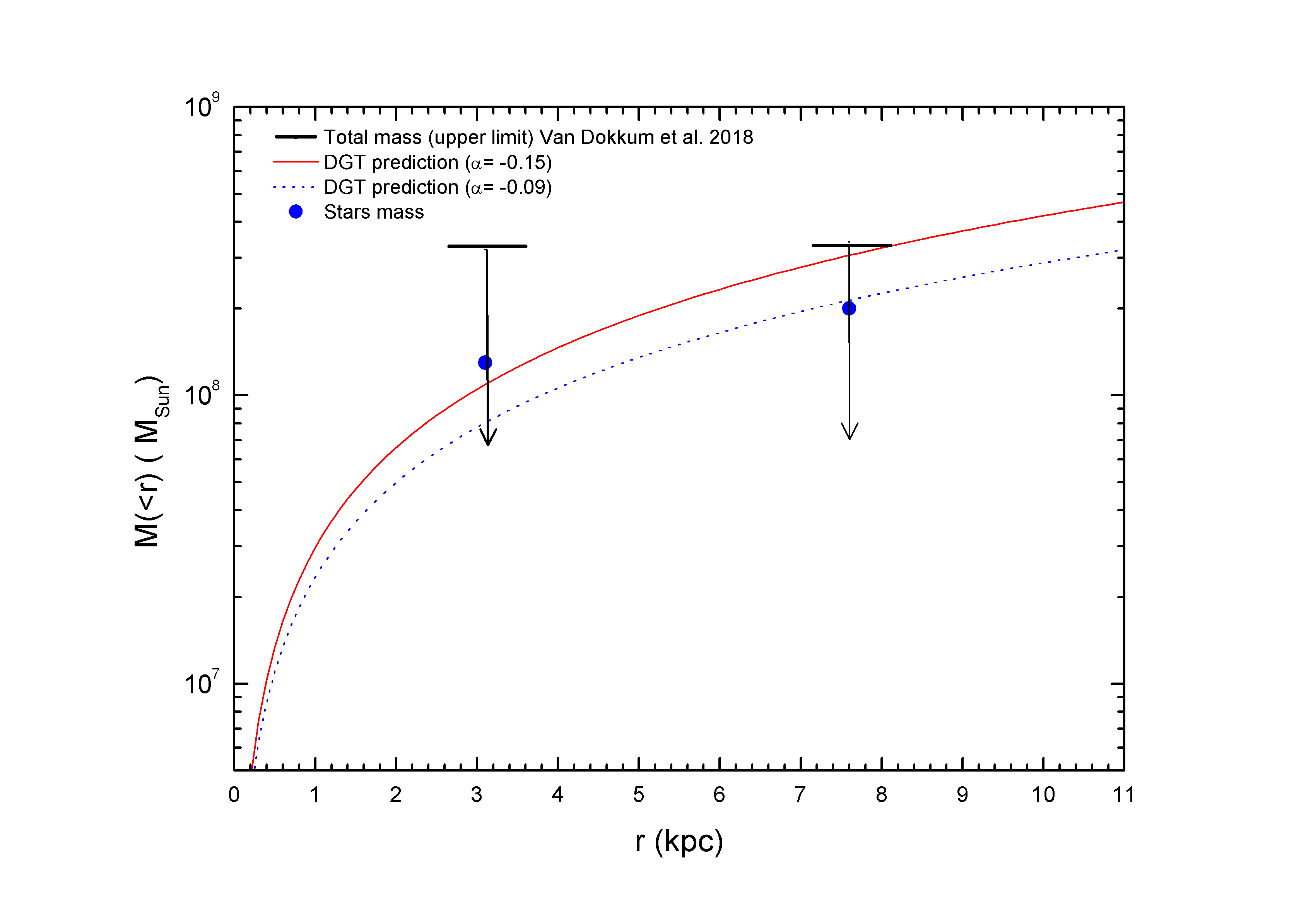}
\vspace*{-0.0cm}
\caption{The mass profiles according to Generalized Faber-Jackson relation (Eq.~\ref{Eq:mass}) using the the NGC1052-DF2 galaxy parameters. The red solid curve and the blue dot curve are the predictions, for $\alpha=-0.15$ (T=8.38K) and $\alpha=-0.09$ (T=10.92 K) respectively. Both are within 90\% upper limits of the total observed mass of NGC1052–DF2.
}
\vspace*{0.0cm}
\label{m_r}
\end{figure}

From the mass profiles of NGC1052-DF2 galaxy reported by \cite{vand18} we can infer
that the enclosed mass $M(<r)$ increases as the radius increases, reaching asymptotically,
the 90\% upper limits on the total mass of NGC1052–DF2 and that is reproduced in Fig.~\ref{m_r}. For a stellar velocity dispersion of $\sigma_{stars} = 8.4 \pm 2.1$ 
km s$^{-1}$ \citep{dani19}, the mass profiles according to Generalized Faber-Jackson relation of DGT (Eq.~\ref{Eq:mass}) 
is within the 90\% upper limits on the total mass of NGC1052–DF2 
for $\alpha$ values between  $\alpha=-0.09$ and $\alpha=-0.15$ as is shown in 
Fig.~\ref{m_r}, blue and red curves respectively.  Thus, according to DGT, the thermal bath temperature of NGC1052–DF2 galaxy is between 8.38 K and 10.92 K. These values are in the $\alpha$-Temperature correlation and are shown in Fig.~\ref{alpha_a}.

Considering that the host galaxy NGC1052 is at a distance of 20 Mpc (z $\sim$ 0.022) from the solar system, its thermal bath temperature is about $T=2.73(1+z)=2.79$ K, so the dwarf galaxy NGC1052-DF2 orbiting this galaxy is subject to additional radiation from it. Thus the thermal bath temperature of  NGC1052-DF2 is estimated from 5.54 K to 8.13 K above 2.79 K.

Following Fig.~\ref{alpha_a} we can see that according to DGT, the only substantial difference between the local dwarf galaxies, orbiting the Milky-Way and the NGC1052-DF2 dwarf galaxy orbiting the NGC1952 galaxy, is the thermal bath temperature. The thermal bath temperature of local dwarf galaxies is up to a degree higher than 
2.73 K \citep{navi18b}. Whereas, the thermal bath temperature of NGC1052-DF2 dwarf galaxy is about 6.5 K higher than the temperature of the local dwarf galaxies. 

The high bath thermal temperature of NGC1052-DF2 relative to local dwarf galaxies can be explained considering that
NGC 1052 is a giant elliptical galaxy, is catalog as LINER-type active galactic nucleus (AGN) which signals the intense starburst activity in the galaxy's center \citep{mich05}. 

Besides, if we consider that the NGC1052-DF2 mass is within a radius of 7.6 kpc
\citep{vand18}, we obtain the correlation of the velocity dispersion and mass. According to
Eq.~\ref{Eq:mass}, this correlation can be written as

\begin{equation}
\sigma=\left(\frac{Ga^{\alpha}}{r^{1-\alpha}}\right)^{1/(2\alpha+2)}M^{1/(2\alpha+2)}
\label{Eq:sigma}
\end{equation}

\begin{figure}
\vspace*{-0.0cm}
\hspace*{0.0cm}
\centering
\includegraphics[width=13.0cm]{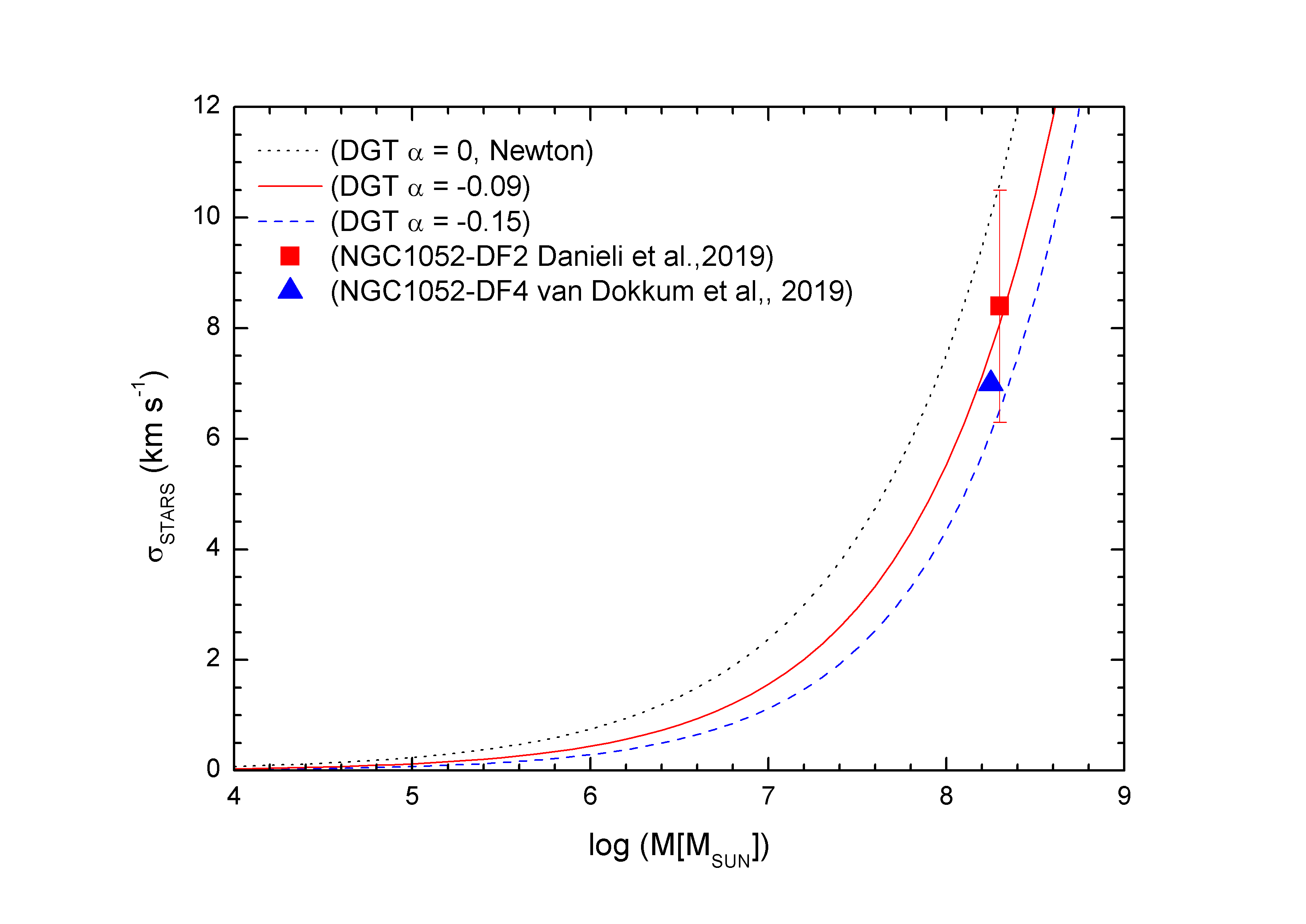}
\vspace*{-0.0cm}
\caption{The relation between stellar mass and velocity dispersion according to DGT prediction
(Eq.~\ref{Eq:sigma}), for three values of $\alpha$. The red square and blue triangle represent observations from \cite{dani19} and \cite{vand19}, respectively.
}
\vspace*{0.0cm}
\label{dispersion}
\end{figure}

Fig.~\ref{dispersion} shows this correlation for three different values of $\alpha$, $\alpha=0$ (Newtonian-regime in DGT), $\alpha$=-0.09 and $\alpha=-0.15$. For comparison we have include
the  KCWI stellar  velocity dispersion  $\sigma_{stars}=8.4 \pm 2.1$ for NGC1052=DF2 \citep{dani19}. We also include the stellar  velocity dispersion  $\sigma_{stars}=7.0$ for NGC1052=DF4 \citep{vand19}. For the NGC1052-DF4 case,
 we assume that this galaxy has approximately the same mass and size than the NGC1052-DF2. Indeed, according to \cite{vand19},
  NGC1052-D4 is in the same group as NGC1052-DF2 and has a similar size, luminosity, morphology and globular cluster population.
 
\section{Conclusions}

Following Fig.~\ref{dispersion}, we can see the predictions of DGT are in agreement with the data. However, a gravitational model to be robust, in addition to correlating the already known data, has to make predictions.
From the structure of DGT to explain the anomaly of the galaxy DGC 1052-DF2 (with an apparent lack of dark matter) it is possible to make at least two predictions:

(a) Several or maybe all of the satellite galaxies of NGC1052 must have
the same behavior of the galaxy NGC1052-DF2, since they must be immersed in thermal baths with almost the same temperature. This behavior was already observed in a second galaxy, the NGC1052-DF4 dwarf galaxy, and that is in the same group of the galaxy NGC1052-DF2 \citep{vand19}.

(b) In general, AGNs are strong emitters of X-rays, ultraviolet, optical, and radio,  and some such as the AGN 421, emitting gamma radiation. If these galaxies had satellites (dwarf) galaxies they must be apparently  like galaxies with lack dark matter, because they must be immersed in thermal baths with temperature above 8.0 K and according to DGT their motions must be close with the Newtonian regime.

The confirmation of these two DGT predictions, about the satellite galaxies of AGNs, will show that we are on the right track.

This work is supported by the Conselho Nacional de Desenvolvimento Cient\'{i}fico e Tecnol\'{o}lgico (CNPq, Brazil, grant 152050/2016-7.

\end{document}